# A Fuzzy Approach to Record Linkages

Pratik K. Biswas

*Abstract*—Record Linkage is the process of identifying and unifying records from various independent data sources. Existing strategies, which can be either *deterministic* or *probabilistic*, often fail to link records satisfactorily under uncertainty. This paper describes an indigenously developed fuzzy linkage method, based on *fuzzy set* techniques, which can effectively account for this *uncertainty* prevalent in the disparate data sources and address the shortcomings of the existing approaches. Extensive testing, evaluation and comparisons have demonstrated the efficacy of this fuzzy approach for record linkages.

## I. Introduction

Data comes in different shapes, sizes, and quality from multiple different sources with many idiosyncrasies. Personal information, encapsulated as strings, often contains typographical and data entry errors and misspellings, over time changes, false, or missing information. Spouses and/or children of a family can use the primary member's information. Households may not have been identified. Therefore, there is a pressing need in every organization to unify, align and merge data gathered from multiple disparate data sources.

Record linkage [1-3] is the task of linking or finding data from multiple data sources that refer to the same entity. There are two main types of linkage algorithms, namely, *deterministic* and *probabilistic*. Choosing the best algorithm to use in a given situation depends on many interacting factors, such as time, resources, the research question, and the quantity and quality of the variables or columns available to link, including the degree to which they individually and collectively are able to identify an individual uniquely. The key is to develop algorithms to extract and make use of enough meaningful information to arrive at sound decisions. Here, we consider record linkages based only on columns whose values are *strings (textual)*.

*Deterministic* algorithms determine if record pairs agree or disagree on a given set of identifiers (attributes) through *exact* pattern matching, where agreement on a given identifier is assessed as a discrete—*all-or-nothing*—outcome. Match status is determined in a single or in multiple steps. Accordingly, there are two types, namely, *exact* and *approximate* or *iterative* record matching. Exact deterministic matching is a single step strategy that requires an exact match on the full set of identifiers all at once. Approximate or iterative matching is a multi-step strategy that requires an exact match on one of the several rounds of matching but not on all possible identifiers. In this strategy, records are matched in a series of progressively less restrictive steps where pairs that do not meet a first round of match criteria are passed to a next round. If a record pair meets the criteria at any step, it is classified as a match. Otherwise, it is a non-match. The iterative deterministic strategy may consist of a sequence of deterministic matches using different match criteria in each successive round.

*Probabilistic* algorithms determine the *likelihood* (*probability*) of agreement between record pairs on a given set of identifiers through *inexact* (*approximate or fuzzy*) string/pattern matching (attribute values). Comparison space can be reduced by searching only those matched pairs that meet certain basic criteria – this is known as *blocking*. Probabilistic strategies look at the discriminatory power of each identifier. The weight assigned to agreement or disagreement on each identifier is assessed as a *likelihood ratio*, comparing the *m*- and *u*- *probabilities* [2], where *m-probability* estimates the probability that two records *agree* on a particular identifier when they are *true matches* while the *u-probability* estimates the probability that two records *agree* on a particular identifier merely *by chance*. When two records agree on an identifier, an *agreement weight* is calculated by dividing the *m-probability* by the *u-probability* and taking the $\log_2$ of the quotient. When two records disagree on an identifier, a *disagreement weight* is calculated by dividing (1 - *m-probability*) by (1 - *u-probability*) and taking the $\log_2$ of the quotient. The *total linkage weight* for each record pair is obtained by summing the individual linkage weights for each linked identifier (column). The total linkage weights of each record pair is then compared with a pre-defined *probability threshold* and those record pairs whose weights are above this threshold (standard = 0.5) are considered as *linkages*.

In information-poor yet business critical scenarios, where the collected data is of inferior quality, deterministic linkages may result in too few *True Positives*, while probabilistic linkages may result in very high *False Positives* or *False Negatives* depending on the chosen *cut-off* probability threshold.

The quantification of uncertainties can be based on either *probability theories*, using *random* variables, or *fuzzy set theories*, using *interval* and *fuzzy* variables. *Fuzzy set*-based techniques are increasingly being used for design problems where there is random uncertainty, but information on the statistics of the uncertainty is sketchy. Fuzzy models are for *uncertainties* due to *vagueness* and can often perform better if little information is available.

In this paper, we present a novel and flexible *fuzzy record linkage* solution that combines *fuzzy matching*, *fuzzy set theory, fuzzy logic*, *fuzzy constraint*, *fuzzy weights*, *fuzzy rule-based inferencing* and *fuzzy clustering* to provide record linkages that are more purposeful than either deterministic or probabilistic solutions in uncertain scenarios with imprecise and inadequate information.

Pratik K. Biswas is with Artificial Intelligence and Data Science, Verizon Communications, Basking Ridge, NJ, USA 07920 (e-mail: Pratik.Biswas@verizonwireless.com).

## II. RELATED WORK

Related research falls into two categories: 1) record linkages and 2) fuzzy techniques associated with the proposed method.

Herzog et al. [1] reviewed the different record linkage techniques. Fellegi and Sunter [2] provided the theory of probabilistic linkages. Christen [3] represented the record linkage procedure as a workflow, whose steps were cleaning, indexing, comparing, classifying and evaluation. The classified record pairs could be fed back to improve the previous step.

Zadeh [4] proposed the fuzzy set theory in 1965 in order to model the uncertain and vague situations more precisely. Dubois and Prade [5] defined the concept of a fuzzy number and Novak [6] alluded to the triangular fuzzy number as the most used type of fuzzy number. Ali et al. [7] described many arithmetic operations on the triangular fuzzy number. Yue et al. [8] extended the traditional definition of neighborhood coverings to fuzzy ones, consisting of membership functions that formed an approximate distribution of neighborhood belongingness.

Zadeh [4] also introduced the notion of fuzzy logic along with his proposal of fuzzy set theory. Furthermore, he [9-11] developed the concept of linguistic variables, which was subsequently formalized by Zimmerman [12]. Borovička [13] described the transformation process of linguistic terms to fuzzy numbers. Chen and Hwang [14] discussed the conversion scale. Wang et al. [15] also showed how to calculate fuzzy weights from linguistic terms. Bellman [16] described the fuzzy decision process as a multi-objective programming problem involving fuzzy goals and fuzzy constraints. Ruttkay [17] considered fuzzy constraint as a fuzzy relation. Mamdani [18] developed a new method of fuzzy inference systems, known as Mamdani-Type Fuzzy Inference. Beginning in 1987, various solutions to the Fuzzy Weighted Average (FWA) problem have been provided. Dong and Wong [19] presented the first FWA algorithm; Liou and Wang [20] presented an improved FWA (IFWA) algorithm; and Lee and Park [21] presented an EFWA. All three algorithms were based on α-cuts. Other solutions included those in [22-25]. Ahmed et al. [26] compared and analyzed the performances of the different algorithms for Fuzzy Analytic Hierarchy Process (FAHP). Bezdek [27] introduced Fuzzy c-means clustering, while Li and Lewis [28] reviewed the application of various fuzzy clustering algorithms.

Finally, Abril et al. [29] introduced a distance-based record linkage method which used Choquet integral to compute the distance between records as well as supervised learning to determine the optimal fuzzy measure for the linkage.

## III. PRELIMINARIES – CONCEPTS AND TERMINOLOGIES

In this section, we clarify key concepts and terminologies and explain certain technologies, which provide the foundation for our work.

### A. Fuzzy Set Theory

In this subsection, we will review only those concepts of fuzzy set theory [4-5, 7] that are relevant for us.

Let $X$ denotes the classical set of objects called the *universe*, the generic elements of which are marked as $x$. A **fuzzy set** $A$ is defined as: $A = \{(x, \mu_A(x)) \mid x \in X, \mu_A(x) \in [0,1]\}$. In the pair $(x, \mu_A(x))$, the first element belongs to the classical set and the second element $\mu_A(x)$, called *membership function*, belongs to the interval $[0,1]$. We can say that $A$ is a subset of $X$ that has no sharp boundaries.

The **kernel** of a fuzzy set $A$ is defined as $Ker(A) = \{x \mid \mu_A(x) = 1\}$, while the **support** of fuzzy set $A$ is defined as: $Sup(A) = \{x \mid \mu_A(x) > 0\}$.

Given $\alpha \in [0,1]$, the **α-cut** of a fuzzy set $A$ is defined as a crisp set $A_\alpha (or\ A[\alpha]) = \{x \mid \mu_A(x) \geq \alpha]\}$.

A **fuzzy number** [5] is a fuzzy set $A$ on $\mathbb{R}$ that must possess at least the following three properties, (i) $A$ must be a normal fuzzy set; (ii) $A_\alpha$ must be closed interval for every $\alpha \in [0,1]$, (iii) the support of $A$, $A_{0+}$, must be bounded. A fuzzy number intuitively represents a value, which is inaccurate.

A **triangular fuzzy number** [6] represented with three points as follows: $A = (a, b, c), where\ a \leq b \leq c$. Its *membership function* has the shape of a triangle, formalized as:

$$\mu_A(x) = \begin{cases} 0, & x < a, x > c \\ \frac{x-a}{b-a}, & a \leq x \leq b \\ \frac{c-x}{c-b}, & b \leq x \leq c \\ 1, & x = b \end{cases} \quad (3.1)$$

The **left** and **right triangular fuzzy number** can be formally written as $A_l = (a, b, b)$, and $A_r = (b, b, c)$, with membership functions [30] as follows:

$$\mu_{A_l}(x) = \begin{cases} 0, & x < a, x > p \\ \frac{x-a}{b-a}, & a \leq x \leq b \\ 1, & b \leq x \leq p \end{cases} \quad (3.2)$$

$$\mu_{A_r}(x) = \begin{cases} 0, & x < 0, x > c \\ \frac{c-x}{c-b}, & b \leq x \leq c \\ 1, & 0 \leq x \leq b \end{cases} \quad (3.3)$$

The **α-cut** of a triangular fuzzy number A = (a, b, c) is defined as $A_a (or\ A[\alpha]) = [a + (b-a)\alpha, c - (c-b)\alpha]$ (3.4)

A **fuzzy neighborhood** of $x \in X$ is defined as
$FN(x) = \{(y, \mu_{FN(x)}(y)) \mid y \in X, \forall x, x \in X, d(x,y) < \lambda, \mu_{FN(x)}(y) \in [0,1] \ \& \ \mu_{FN(x)}(y) = f(d(x,y)) \}$, where $d(\bullet)$ is distance function and $\lambda$ is the threshold. $f(d(x,y))$ can be $(1 - d(x,y)/\max(d(x,y)))$.

Further, $FN_X = \{FN(x_1), FN(x_2), \ldots, FN(x_n)\}$, i.e., the set of fuzzy neighborhoods of data samples forms the **fuzzy neighborhood covering** of the data space $X$.

A binary **fuzzy relation** R between crisp sets $X$ and $Y$ is a set of ordered pairs $\{((x, y), \mu_R(x, y)) \mid x \in X, y \in Y, R \subseteq X \times Y, \mu_R(x,y) \in [0,1]\}$. The membership function indicates the intensity of the relation between $x$ and $y$.

A **fuzzy constraint** $C$ is a n-ary fuzzy relational mapping from product of finite domain of variables to the [0,1] interval and is defined as:
$C = \{(\langle x_1, x_2, \cdots, x_n \rangle, \mu_C(\langle x_1, x_2, \cdots, x_n \rangle)) \mid \langle x_1, x_2, \cdots, x_n \rangle \in D_1 \times D_2 \times \cdots \times D_n,$
$\mu_C(\langle x_1, x_2, \cdots, x_n \rangle) \in [0,1]\}$ (3.5)

We call the membership function $\mu_C(\langle x_1, x_2, \cdots, x_n \rangle)$ the *degree of satisfaction* of the constraint $C$. If $\mu_C(\langle x_1, x_2, \cdots, x_n \rangle) = 1$, then we say that $\langle x_1, x_2, \cdots, x_n \rangle$ *fully satisfies* $C$ (*crisp constraint*). If $\mu_C(\langle x_1, x_2, \cdots, x_n \rangle) = 0$, then we say that $\langle x_1, x_2, \cdots$ $, x_n \rangle$ *fully violates* $C$. We define the *support set* of $C$ as $Sup(C) = \{\langle x_1, x_2, \cdots, x_n \rangle \mid \mu_C(\langle x_1, x_2, \cdots, x_n \rangle) > 0\}$. For elements which satisfy the constraint better than a given threshold $\alpha$, where $1 \geq \alpha > 0$. The $\alpha - support$ set is defined as follows: $Sup(C)_\alpha$ (or $Sup(C)[\alpha]$) = $\{\langle x_1, x_2, \cdots, x_n \rangle \mid \mu_C(\langle x_1, x_2, \cdots, x_n \rangle) \geq \alpha\}$.

### B. Fuzzy Logic, Linguistic Variables and Fuzzy System

**Fuzzy logic** [4] may be defined as *many-valued logic* in which the *truth value* of variables may be expressed through *linguistic terms* and may be any real number between 0 and 1. It is employed to handle the concept of *partial truth*, where the truth value may range between *completely true* and *completely false*. Fuzzy logic is an approach to computing based on *degrees of truth* rather than on *true* or *false* (1 or 0) as in Boolean Logic. Fuzzy logic is based on the observation that people make decisions based on imprecise and non-numerical information.

A **fuzzy or linguistic variable** [4, 9-11, 13, 15] is a variable that can accept non-numeric values, i.e., linguistic terms to facilitate expression of facts and rules. Because natural languages do not always contain enough value terms to express a fuzzy value scale [14], it is common practice to modify linguistic values with adjectives or adverbs.

In general, a linguistic variable has values that are words and the meanings of these words are fuzzy sets in a certain universe. The values of a linguistic variable are called *terms*. Typical examples of this type of variable could be age, height, intelligence, etc. Their values could be terms like young, high, clever, etc. Each term can be represented by a *triangular* (or *trapezoidal*) fuzzy number.

A **fuzzy system** [18] is a repository of fuzzy expert knowledge that can reason data in vague terms instead of precise Boolean logic. The expert knowledge is a collection of fuzzy membership functions and **fuzzy rules,** used to infer an output based on input variables. Unlike in crisp logic, the antecedent and the consequence of a fuzzy rule can be true to a degree, instead of being entirely true or entirely false. **Fuzzy inference** can be viewed as decision making with a fuzzy system employing fuzzy implications and fuzzy rules of inference. **Defuzzification** converts fuzzy values into crisp quantities, that is, it links a single point to a fuzzy set, given that the point belongs to the support of the fuzzy set.

### C. Fuzzy Analytic Hierarch Process

**Fuzzy Analytic Hierarchy Process** (FAHP) [26, 31] is a method that combines Analytic Hierarchy Process (AHP) with fuzzy logic theory. FAHP is a multi-criterion technique that works similarly as the AHP method, but scales into a triangular fuzzy number scale for representing the criteria relevance. FAHP can be used to infer *fuzzy weights* from linguistic terms for fuzzy (linguistic) variables using either *geometric mean* method [30] or *extent analysis* method or *entropy-based* method. These fuzzy weights can be further defuzzified.

### D. Fuzzy Weighted Average

We have defined the **Fuzzy Weighted Average** as in [22] by applying Zadeh's extension principle with α-cut to the non-fuzzy weighted average ($wa$). Let $X_1, X_2, \cdots, X_n$ (variables), and $W_1, W_2, \cdots, W_n$ (weights) be the fuzzy numbers, then the **fuzzy weighted average** ($fwa$) is an *interval* defined as:
$fwa(X_1, X_2, \cdots, X_n, W, W_2, \cdots, W_n)[\alpha], 1 \leq i \leq n$
$= \{wa(x_1, x_2, \cdots, x_n, w_1, w_2, \cdots, w_n) \mid x_i \in X_i[\alpha], w_i \in W_i[\alpha]\}$
$= \{\sum_{i=1}^n x_i * w_i \mid x_i \in X_i[\alpha], w_i \in W_i[\alpha]; \sum_{i=1}^n w_i = 1\}$ (3.6)
We will consider weights which are triangular fuzzy numbers, i.e., $W_i = (l_i, m_i, r_i)$. The condition of fwa-normality is $\sum_{i=1}^n m_i = 1$, i.e., their modal values add up to unity. When a set of fuzzy weights is not fwa-normal, they can be made normal by normalizing it. We have used the algorithm in [22] to compute the fuzzy weighted average.

### E. Fuzzy C-means Clustering

**Fuzzy c-means clustering (FCM)** [27, 32] is a *soft clustering* [28] approach, where each data point is assigned a membership score that indicates the degree to which it belongs to a cluster. The main advantage of fuzzy c-means clustering is that it allows gradual memberships of data points to clusters measured as degrees in [0,1]. This gives the flexibility to express overlapping membership, i.e., that data points can belong to more than one cluster.

### F. Fuzzy Matching

**Fuzzy matching**, also called *Approximate String Matching*, is a technique that helps identify two elements of text, strings, or entries that are approximately similar but are not exactly the same. It is called *fuzzy* because of the *inexact* nature of the matching involved. The *closeness* of a match, known as *edit distance*, is measured in terms of the number of primitive operations necessary to convert the string into an exact match.

## IV. FUZZY RECORD LINKAGE

This section provides a description of a *fuzzy linkage technique* that we are proposing for linking database records. This record linkage technique **uniquely** integrates *approximate (fuzzy) string/pattern matching, fuzzy logic, search space reduction through fuzzy constraints, fuzzy weighted linkage estimation using fuzzy (linguistic) weights, fuzzy rule-based inferences and fuzzy clustering* through a 5-step sequential procedure/method.

The procedure is highly parameterized. The list of parameters to the proposed procedure includes: *Constraint Type, Field & Value* (default: "None/False"), *Link Columns*, *Match Algorithms, Logic Type* (default: "Fuzzy"), *Thresholds* (default: 0.9), *Linkage Type* (default: "crisp"), *Crisp Weight Vector* (default: <1,1,1>), *Linguistic Terms* (default = ("low", "medium", "high")), *Fuzzy Number Scale*, *FAHP Method* (default: "geometric mean"), *Fuzzy Rule Base, No. of Fuzzy Clusters* (default: 3). Parameters will be italicized in text.

The procedure supports "crisp" & "fuzzy" *Constraint Types* "Boolean" & "Fuzzy" *Logic Types* and "crisp" & "fuzzy" *Linkage Types* for record linkages. "Boolean" *Logic Type* is associated only with "crisp" *Linkage Type* and *Crisp Weight Vector* (or default weights of 1 for the *Link Columns*) to establish record linkages. "Fuzzy" *Logic Type* is associated with either "crisp" or "fuzzy" *Linkage Type* and FAHP generated crisp (ignoring the *Crisp Weight Vector*) or fuzzy weights for the *Link Columns* to establish record linkages. FAHP is used to generate fuzzy weights from *Linguistic Terms* describing the relevance of the *Link Columns*, i.e., the columns to be linked in a record pair. Fuzzy rules from inputted fuzzy rule base connect match scores of the linked columns to the total linkage score for the record pair by treating them all as fuzzy variables. "Fuzzy" *Logic Type* and "fuzzy" *Linkage Type* employ Mamdani type [18] fuzzy inference triggering fuzzy rules. The objective of the procedure is two-fold. First, to associate a total linkage score to each linked record pair indicating the quality of the linkage so that it can be ranked individually if needed, and second, to cluster the linked record pairs into a pre-specified number of fuzzy clusters based on their linkage qualities so that the different clusters can be interpreted and treated differently as per user requirements. Fig. 1 shows the steps of this procedure.

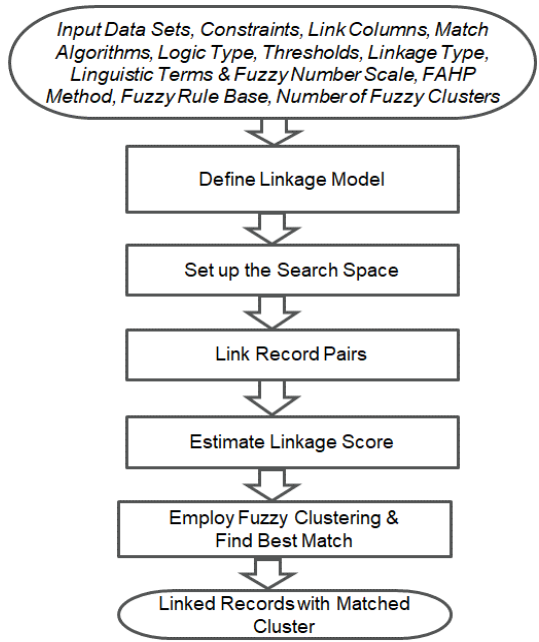

Fig.1: Proposed Fuzzy Record Linkage Solution

Next, we take a look at all the key steps of this *Fuzzy Record Linkage (FRL)* procedure and discuss their implementations.

### A. Linkage Model

This is the first step of the procedure where it sets up the linkage query using the inputted parameter values. It consists of the following two activities:
i. Define the linkage model with the parameters to the procedure, i.e., *Link Columns*, *Match Algorithms*, *Logic Type*, *Thresholds*, *Linkage Type*, etc.
ii. Construct a *linkage query* with the linkage model using the inputted values to the aforesaid parameters

The result provides definition and invocation of the linkage model. A linkage query can support either "Boolean" or "Fuzzy" *Logic Types* and "crisp" or "fuzzy" *Linkage Types*.

### B. Search Space

This step sets up the constrained search space using the input data sets and the constraints (if any). The constraint can be *crisp* (exact) or *fuzzy*. This step consists of the following activities:
i. Merge the input data sets. With 2 data sets, e.g., $A$ & $B$ of sizes: $m$ & $n$; the search space is $A \times B$ and has $m * n$ (size) record pairs
ii. *Constrain* the search space $\{(x, y) \mid x \in A \ \& \ y \in B\}$ using *blocking* with either *crisp* ($\mu_C(\langle x, y \rangle) = 1$) or *fuzzy constraint*, i.e., its $\alpha - support$ set with $\mu_C(\langle x, y \rangle) \geq \alpha$, defined through either *crisp* (exact) or *fuzzy neighborhoods* (i.e., fuzzy neighborhood covering), whose memberships ($\mu_{FN(x)}(y)$) can be determined using either the *edit distances* or physical distances (in case of locations)

The result is a merged data set for a constrained search space consisting of paired records.

### C. Record Pair Linkage

Our focus here is on linking record pairs through the match scores on the linked columns, *whose original values are strings*. So, this step links each record pair in the constrained search space by applying the selected fuzzy (approximate) string *Matching Algorithm* on the *Link Columns* (columns to be linked) to generate the corresponding match scores in [0, 1] and then transforming them if needed depending on the chosen *Logic Type*. It consists of the following two activities:
i. Execute the linkage query by applying *fuzzy (approximate) string matching* on the *Link Columns* to generate the respective *match scores*
ii. If *Logic Type* is "Boolean" then
   a. For every *Link Column* in a record pair, set *match score* to 1, if match score ≥ match *Threshold*; otherwise set it to 0
   b. Ensure that *Linkage Type* is set to "crisp"

The result is the constrained data set from step A, with each linked record pair now additionally including either replaced (1 or 0) or raw *match scores* in [0, 1] for each *Link Column*.

### D. Linkage Score Estimation

Linkage score estimation is the most important step of this procedure. *Total linkage score* (*TS*) reflects the quality, i.e., strength of record linkage. This step computes the total linkage score for each linked record pair from the match scores and the weights associated with the *Link Columns*, based on the *Logic and Linkage Types*. If "Boolean" *Logic Type* is chosen then *Linkage type* can only be "crisp", and normalized *Crisp Weight Vector* (or default weight of 1 for each *Link Column*) is used to compute the total linkage score of each record pair. FAHP is used only with "Fuzzy" *Logic Type* for record linkages. It is used to estimate fuzzy weights (triangular fuzzy numbers) from the specified *Linguistic Terms*, reflecting the relative importance of the columns to be linked, and the inputted *Number Scale* (both parameters to the procedure). This relative importance of the *Link Columns* can come from either the customers (users) or the owner of the data sources based on their experience and understanding of the data usage. Furthermore, if the chosen *Linkage Type* is "crisp" then the fuzzy weights for the *Link Columns* are defuzzified and normalized (ignoring the *Crisp weight Vector*) for computing the weighted average of the match scores reflecting the total linkage score for the record pair. For example, if two data sets are to be linked on three columns: Name, Address, and City and their deemed relevancies are "low", "medium" and "high", then FAHP with a 3-point number scale can yield their fuzzy weights as (0.1, 0.16, 0.33), (0.15, 0.3, 0.6), (0.28, 0.54, 0.96), which can be further defuzzified, if needed, into 0.17, 0.31, and 0.52 to serve as their normalized crisp weights. However, if the *Linkage Type* is "fuzzy" then fuzzy weights (from FAHP) and match score intervals (represented as triangular fuzzy numbers) of the *Link Columns* are used to estimate the fuzzy weighted average to represent the interval of the total linkage score of the linked record pair, also as a triangular fuzzy number. The match scores of the *Link Columns* and the total linkage score of the linked record pair become fuzzy variables whose values can be *Linguistic Terms*, which are fuzzy sets represented by triangular fuzzy numbers based on the intervals and distributions of their original values. Fuzzy rules are used to connect the fuzzy variables, i.e., fuzzy match scores of the *Link Columns* in the antecedent to the fuzzy weighted average (reflecting the total linkage score) in the consequence. A Mamdani type [18] fuzzy controller applying fuzzy inference on these fuzzy rules can then infer a defuzzified crisp total linkage score for each linked record pair from the raw match scores (of the *Link Columns*). In this context, we establish a formal framework for the two types of record linkages supported by the proposed procedure.

**Definition 4.1:** *A crisp linkage is a record linkage whose quality is estimated as a weighted average (wa) of match scores of the linked columns, where the match scores and weights are both crisp numbers.*

**Definition 4.2:** *A fuzzy linkage is a record linkage whose quality is estimated as a fuzzy weighted average (fwa) of match scores of the linked columns, where the match scores and weights are both fuzzy numbers.*

Let $S = \langle s_1, s_2, \cdots, s_n \rangle$ and $W = \langle w_1, w_2, \cdots, w_n \rangle$ be the match scores ($[0,1]$) and normalized crisp weights of the *Link Columns* $\langle x_1, x_2, \cdots, x_n \rangle$.

**Proposition 4.1:** *Crisp linkage is a mapping from X ($\langle x_1, x_2, \cdots, x_n \rangle$) to $\mathbb{R}$ that results in a crisp relation defined as:*
$CL = \{(\langle s_1, s_2, \cdots, s_n \rangle, TS) \mid TS \, \varepsilon \, \mathbb{R} \text{ and } TS = wa(s_1, s_2, \cdots, s_n, w_1, w_2, \cdots, w_n) = S \bullet W \}$.

Let $\langle FS_1, FS_2, \cdots, FS_n \rangle$ and $\langle FW_1, FW_2, \cdots, FW_n \rangle$ be the fuzzy match scores & fuzzy weights of the *Link Columns* $\langle x_1, x_2, \cdots, x_n \rangle$.

**Proposition 4.2:** *Fuzzy linkage is a mapping from X ($\langle x_1, x_2, \cdots, x_n \rangle$) $\times$ $\mathbb{R}$ to $[0,1]$ that results in a fuzzy relation defined as:*
$FL = \{((\langle s_1, s_2, \cdots, s_n \rangle, TS), \mu_{FL}(\langle s_1, s_2, \cdots, s_n \rangle, TS)) \mid \mu_{FL}(\langle s_1, s, \cdots, s_n \rangle, TS) \in [0,1], \mu_{FL}(\langle s_1, s_2, \cdots, s_n \rangle, TS) = \mu_{FL}(TS); TS = fwa(FS_1, FS_2, \cdots, FS_n, FW_1, FW_2, \cdots, FW_n)[\alpha], \alpha \in [0,1]\}$.

Let $S_{CL}$ & $S_{FL}$ represent the intervals of total linkage scores ($TS$) from crisp and fuzzy linkages, and let $\subset$ denotes sub-interval, then the following can be remarked about the intervals in our context.

**Remark 4.1:** $S_{CL} \subseteq [0, 1]$

*Proof*: The total linkage score for *crisp linkage* is estimated as the weighted average of match scores ($s_i$) and normalized crisp weights ($w_i$) of the *Link Columns*, i.e., $\sum_{i=1}^{n} s_i * w_i$. Since $0 \leq s_i \leq 1$, and $\sum_{i=1}^{n} w_i = 1$, we have $0 \leq \sum_{i=1}^{n} s_i * w_i \leq 1$. This implies that $S_{CL} \subseteq [0\ 1]$.

**Remark 4.2**: $S_{FL} \subseteq [0, 1]$

*Proof*: The total linkage score for *fuzzy linkage* is estimated as the fuzzy weighted average ($fwa$) of fuzzy match scores ($FS_i$) and fuzzy weights ($FW_i$) of the *Link Columns*. The $\alpha$-cuts of $FS_i$ & $FW_i$ are intervals and let them be denoted by $[FS_{i\alpha-}, FS_{i\alpha+}]$ & $[FW_{i\alpha-}, FW_{i\alpha+}]$. From Eqn. (3.6), for $0 \leq \alpha \leq 1$, $fwa(FS_1, FS_2, \cdots, FS_n, FW_1, FW_2, \cdots, FW_n)[\alpha] = \{\sum_{i=1}^{n} fs_i * fw_i \mid fs_i \in [FS_{i\alpha-}, FS_{i\alpha+}], fw_i \in [FW_{i\alpha-}, FW_{i\alpha+}]; \sum_{i=1}^{n} fw_i = 1\}$ is also an interval. Now, it has been shown in [22] that this interval for $fwa$ can be given by $[\sum_{i=1}^{n} FS_{i\alpha-} * FW'_{i\alpha-}, \sum_{i=1}^{n} FS_{i\alpha+} * FW'_{i\alpha+}]$ where $\forall i$, $FW_{i\alpha-} \leq FW'_{i\alpha-} \leq FW_{i\alpha+}$, $FW_{i\alpha-} \leq FW'_{i\alpha+} \leq FW_{i\alpha+}$ & $FW'_{i\alpha-} \leq FW'_{i\alpha+}$. Now, $FS_{i\alpha-}$ & $FS_{i\alpha+}$ are elements of $\alpha$-cuts of fuzzy set $FS_i$, defined over the interval $[0, 1]$ of $s_i$. So, $\forall i$, $0 \leq FS_{i\alpha-} \leq 1$ and $0 \leq FS_{i\alpha+} \leq 1$. Also, by definition of $fwa$, we have $\sum_{i=1}^{n} FW'_{i\alpha-} = 1$ and $\sum_{i=1}^{n} FW'_{i\alpha+} = 1$. This yields $0 \leq \sum_{i=1}^{n} FS_{i\alpha-} * FW'_{i\alpha-} \leq 1$ and $0 \leq \sum_{i=1}^{n} FS_{i\alpha+} * FW'_{i\alpha+} \leq 1$. So, the interval of $fwa$ is bounded by 0 and 1, which are its endpoints, i.e., $[\sum_{i=1}^{n} FS_{i\alpha-} * FW'_{i\alpha-}, \sum_{i=1}^{n} FS_{i\alpha+} * FW'_{i\alpha+}] \subseteq [0,1]$. This implies $S_{FL} \subseteq [0, 1]$.

From **Remarks 4.1 & 4.2**, we can conclude that $S_{CL} = S_{FL}$.

Finally, this step consists of the following activities:

i. If *Logic Type* is "Fuzzy", use *FAHP* to infer *fuzzy weights* (*triangular fuzzy numbers*) associated with the *Linguistic Terms* (for the column-relevance) using the *Fuzzy Number Scale* and the specified *FAHP Method*

ii. If *Linkage Type* is "crisp" then
  a. If *Logic Type* is "Boolean", then either use the normalized *Crisp Weight Vector* (or default weights of 1 for *Link Columns*); else if *Logic Type* is "Fuzzy", then *defuzzify* the *fuzzy weights* (step i) to generate normalized *crisp weight* for each *Linguistic Term* describing each linked column's relevance/importance
  b. Calculate *total linkage score* for each linked record pair as $TS = S \bullet W$, along with the *weighted score* ($s_i * w_i$) of each *Link Column i*, where $S = <s_1, s_2, \cdots, s_n>$ be *score vector* and $W = <w_1, w_2, \ldots, w_n>$ be the normalized *crisp weight vector* for the *Link Columns*

iii. If *Linkage Type* is "fuzzy" then
  a. Extract the $\alpha$-cuts from the *triangular fuzzy numbers* representing *score intervals (with modes)* and the *fuzzy weights* of the *Link Columns* using Eqn. (3.4)
  b. Compute *total linkage score* (*TS*) as a *fuzzy weighted average* of the $\alpha$-cuts of the *fuzzy scores* and *weights* using Eqn. (3.6) and represent it as a *triangular fuzzy number* (*interval*) [e.g., fuzzy match scores of [(0, 0.25, 1), (0.23, 0.55, 1), (0, 0, 1)] and fuzzy weights of [(0.1, 0.16, 0.33), (0.15, 0.3, 0.6), (0.28, 0.54, 0.96)] for the 3 *Link Columns*: Name, Address, and City of a record pair produce a *fuzzy weighted average* of (0.04, 0.21, 1) using Eqn. (3.6)]
  c. Define *fuzzy sets* for the *Linguistic Terms* using *sub-intervals* from the *intervals* and *distributions* of the *match scores* of the *Link Columns*, as well as the computed *interval* for *TS* (from iii. b) [e.g., Fig. 2 shows 3 fuzzy sets for 3 *Linguistic Terms*: "low", "medium" and "high" covering the full intervals of match scores of the 3 *Link Columns*: Name, Address, City, and *TS* (Score) of the linked record pairs]
  d. Define a Mamdani type [18] *fuzzy controller* [33] using the *fuzzy sets* for the *match scores* & *TS* and the *fuzzy*

*rules* (inputted fuzzy rule-base) connecting the *linked column scores* to the *total linkage score* [e.g., a fuzzy rule based on Fig. 2, using the 3 Link Columns may have the form:

rule1 ← Rule(Name['high'] & Address['high'] & City['low'], Score['medium']), where Name, Address, & City are fuzzy variables, taking values from *Linguistic Terms*: "low", "medium" & "high". Fig. 3a shows the structure of a fuzzy rule, where the nodes represent antecedents, consequent, and terms]

e. Simulate the controller and employ *fuzzy inference* to estimate defuzzified *total linkage score* for each linked record pair from the *match scores* of *Link Columns* [Fig. 3b shows the defuzzified total linkage score, i.e., *TS* (Score) for a record pair]

The result is the data set from the previous step, with each linked record pair additionally including *weighted match scores* for the *Link Columns* (*crisp linkage*) and the estimated *total linkage score* (*TS*).

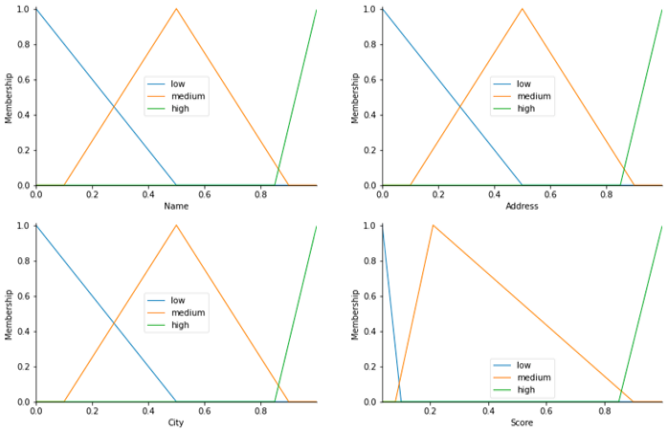

Fig. 2: Fuzzy Sets for Linguistic Terms describing the match scores of *Link Columns* (fuzzy variables: City, Address, Name, & Score)

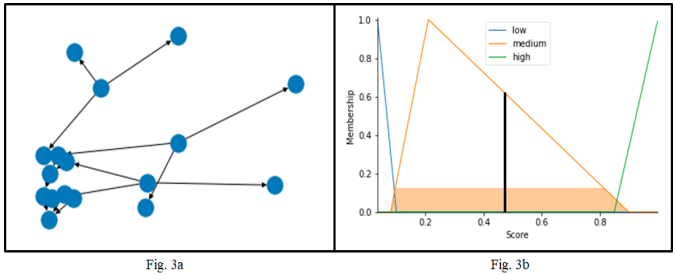

Fig. 3a           Fig. 3b

Fig. 3: A Fuzzy Rule from Fuzzy Rule-base and an estimated Defuzzified Total Linkage Score (*TS*) for a record pair

E. *Fuzzy Clustering*

This is the last step of the main procedure and consists of the following two activities:

i. Use *fuzzy c-means* clustering on the *total linkage scores* from the previous step and generate specified number (*No. of Fuzzy Clusters*) of *overlapping clusters*
ii. For each linked record pair, identify the *cluster* with the *highest membership value* as the *best match*

The result is the data set from the previous step with each linked record pair now including the *best linguistic cluster* to which it belongs. **Algorithm 1** formalizes this entire procedure.

---

**Algorithm 1**
Algorithm for Fuzzy Record Linkages

**Inputs:** Input Sets (A, B), Constraint (C, λ), Link Columns (linkColumns), Match Algorithm (MA), Logic Type (logicType), threshold (th), Linkage Type (linkageType), Linguistic Terms (terms: {"low", "medium", "high"}), Crisp Weight Vector (CW), Fuzzy Number Scale (ns), FAHP-method (FAHP), FWA-α (α), Fuzzy Rule Base (FRB), No. of Fuzzy Clusters (k)

**Output:** Linked record pairs with designated fuzzy cluster

$FRL(C, linkColumns, MA, logicType, linkageType, \lambda, th,$
$\alpha, terms, ns, FAHP, FRB, k)$

$n \leftarrow |linkColumns|$
$x \leftarrow \langle x_1, x_2, \cdots, x_n \rangle$, where $x_i \varepsilon linkCcolumns$
$y \leftarrow \langle y_1, y_2, \cdots, y_n \rangle$, where $y_i \varepsilon linkCcolumns$
$XY \leftarrow \{(x,y) \mid \mu_C(x,y) \geq \lambda, (x,y) \in A \times B, y \in FN(x),$
$\mu_C(x,y) \in [0,1], \mu_{FN(x)}(y) \in [0,1],$
$\mu_C(x,y) \leftarrow \mu_{FN(x)}(y), \mu_{FN(x)}(y)$
$\leftarrow (1 - d(x,y)/d_{max}), d_{max}$
$\leftarrow \max(d(x,y))\}$
$m \leftarrow |XY|$
$S \leftarrow \{\langle s_1, s_2, \cdots, s_n \rangle_j \mid \langle s_1, s_2, \cdots, s_n \rangle \leftarrow MA(x,y), s_i \in [0,1],$
$1 \leq i \leq n, 1 \leq j \leq m\}$

**if** logicType == "Boolean" **then**
  $S \leftarrow \{\langle s_1, s_2, \cdots, s_n \rangle_j \mid s_i \leftarrow (0, if \; s_i \leq th; 1, if \; s_i > th),$
$1 \leq i \leq n, 1 \leq j \leq m\}$
  **if** linkageType ≠ "crisp" **then**
    $linkageType \leftarrow$ "crisp"
  **if** CW **then**
    $TS \leftarrow \{TS_1, TS_2, \cdots, TS_m \mid TS_j \leftarrow S_j \bullet CW', \; S_j \leftarrow$
$\langle s_1, s_2, \cdots, s_n \rangle_j; \; CW' \leftarrow \langle w'_1, w'_2, \cdots, w'_n \rangle, \; w'_i \leftarrow$
$w_i / \sum_{i=1}^{n} w_i, \; CW \leftarrow \langle w, w_2, \cdots, w_n \rangle; 1 \leq i \leq n\}$
  **else**
    $TS \leftarrow \{TS_1, TS_2, \cdots, TS_m \mid TS_j \leftarrow S_j \bullet W, \; S_j$
$\leftarrow \langle s_1, s_2, \cdots, s_n \rangle_j, W \leftarrow \langle 1/n, 1/n, \cdots, 1/n \rangle\}$
**else** (logicType == "Fuzzy")
  $w_i \leftarrow (a, b, c) \leftarrow FAHP(terms, ns), \; a \leq b \leq c, 1 \leq i \leq n$
  **if** linkageType == "crisp" **then**
    $W \leftarrow \langle dw_1, dw_2, \cdots, dw_n \rangle \leftarrow defuzzify(w_i), 1 \leq i \leq n$
    $TS \leftarrow \{TS_1, TS_2, \cdots, TS_m \mid TS_j \leftarrow S_j \bullet W,$
$S_j \leftarrow \langle s_1, s_2, \cdots, s_n \rangle_j\}$
  **else** (linkageType == "fuzzy")
    $s_{i'} \leftarrow (p_i, q_i, r_i), \quad p_i \leq q_i \leq r_i, \; p_i \leftarrow min(s_i),$
$q_i \leftarrow mode(s_i), r_i \leftarrow max(s_i),$
$1 \leq i \leq n,$
$TS \leftarrow (u, v, z) \leftarrow fwa(s_{1'}, s_{2'}, \cdots, s_{n'}, w_1, w_2, \cdots, w_n)[\alpha],$
$u \leq v \leq z$
    **for** $i = 1$ **to** $n$ **do**
      $x_i["low"] \leftarrow (d_i, d_i, e_i), \; d_i < e_i, \; p_i \leq d_i$
      $x_i["medium"] \leftarrow (e_i, f_i, g_i), \quad e_i < f_i < g_i$
      $x_i["high"] \leftarrow (g_i, h_i, h_i), \; g_i < h_i, \; h_i \leq r_i$

    $TS["low"] \leftarrow (l, l, m_1), \; l < m_1, \; u \leq l$
    $TS["medium"] \leftarrow (m_1, m_2, m_3), \; m_1 < m_2 < m_3$
    $TS["high"] \leftarrow (m_3, o, o), \; m_3 < o, \; o \leq z$

    $r_i \leftarrow TS[t] \leftarrow x_1[t] \& x_2[t] \& \cdots \& x_n[t], t \in terms$
    $FRB \leftarrow \{r_1, r_2, \cdots, r_{3^n} \mid 3^n : \text{number of rules}\}$
    $controller \leftarrow fuzzyController(FRB)$
    $linkage \leftarrow simulation(controller)$
    $TS \leftarrow \{TS_1, TS_2, \cdots, TS_m \mid TS_j$
$\leftarrow fuzzyInfer(linkage(S_j)),$
$S_j \leftarrow \langle s_1, s_2, \cdots, s_n \rangle_j, 1 \leq j \leq m\}$
$m \leftarrow FCM(TS, k)$, where $k$ is the number of clusters
$cm \leftarrow argmax(m, axis = 0)$

## V. Performance Evaluation

We have evaluated the performance of the proposed *fuzzy record linkage* method on benchmark US hospital data sets and compared its performance with those of deterministic and probabilistic linkage methods on the same data sets using both *crisp* and *fuzzy constraints* as well as *Boolean* and *Fuzzy logic*.

### A. Experimental Setup

We implemented the *fuzzy record linkage* method in a *python* based *Jupyter Notebook*, using packages like *pandas*, *numpy*, *swifter*, *recordlinkage* and *scikit-fuzzy (skfuzzy)*. For the *fuzzy controller* and the *clustering algorithm*, we used the *control* and the *cluster.cmeans* modules from *skfuzzy*. We tested the proposed method on two publicly-available benchmark data sets of US medical facilities [34], one consisting of **5339** records and the other consisting of **2697** records. The first was an internal data set that contained basic hospital account number, name and ownership information. The second data set contained hospital information (called provider) as well as the number of discharges and Medicare payments for specific Heart Failure procedures. The constrained search space, comprising of either "exact same state" (*crisp constraint*) based record pairs from the two data sets or the *fuzzy neighborhoods* (using edit distances) of the "same state" (*fuzzy constraint*) based record pairs, included **475,830** and **998,860** record pairs respectively. The records from the two data sets were *exactly matched* on City and *approximately* matched (*fuzzy string matching)* on 2 other fields, namely, Hospital Name and Hospital Address using *Levenshtein* and *Jaro-winkler* edit distances. We used both *Boolean* and *Fuzzy logic* as well as *crisp* and *fuzzy linkages*. We considered relevance criteria of "low", "medium" and "high" for Hospital Name, Hospital Address and City respectively, for estimating their *fuzzy weights*. Accordingly for *fuzzy linkages*, we defined 3 *fuzzy sets* for each of the 3 *Link Columns* based on the distribution of their match scores, as shown in Fig. 2. We included a *fuzzy rule-base* associating the three *Link Columns* (fuzzy variables) with a fuzzy score variable and employed a Mamdani type *fuzzy controller* [18, 33] to infer *total linkage score* (*TS*) for each record pair (Fig. 3b). We applied *fuzzy c-means* clustering algorithm on the estimated *total linkage scores* of the linked record pairs and generated 3 overlapping clusters: namely, *Matches*, *Possible Matches* and *Non-matches*. We compared our model performance with those from open-source, off-the-shelf packages like *recordlinkage* (*exact deterministic*) [35] and *splink* (*probabilistic*) [36] only on the *crisply constrained* search space, as these packages did not support *fuzzy* constraints. For non-package-specific strategy comparisons, we used the *fuzzily* constrained search space.

### B. Evaluation Metrics

We evaluated our method for its *effectiveness*. For evaluation of the *effectiveness* of our method, we have considered the number of *Matches* and *Possible Matches* amongst the linked record pairs in the constrained search spaces. We have compared our method with *deterministic* and *probabilistic* linkages on only the number of *Matches*. Note that deterministic linkage results in discrete *all-or-nothing* outcome. Therefore, there are only *Matches* or *Non-matches* but no *Possible Matches*. For probabilistic linkage, we have considered 0.5 (model standard) as cut-off probability threshold and a probability ≥ 0.5 as a *Match*. For us, *Non-matches* were those record pairs that neither *matched* nor *possibly matched*, i.e., (*Total number of linked record pairs – (Matches + Possible Matches*)). They are not shown in Figs. 4 and 5 as their numbers were much larger compared to the other two clusters and were not the focus of our interest.

### C. Discussion on Results and Model Comparisons

Fig. 4 compares the performance of the proposed method with those of *recordlinkage* (for *exact deterministic*) and *splink* (for *probabilistic*) packages on the *crisply* constrained search space consisting of **475,830** linked record pairs. For the proposed solution, we used: i) *Boolean logic* to evaluate the results of approximate string matching on the *Link Columns*, to be consistent with the results from the other packages (any match score ≥ 0.9 was a *Match* and resulted in a score of 1), and ii) *crisp linkage* with both user provided weights as well as the default weight of 1 for each *Link Column* to compute the *total linkage score* (*TS*) for each record pair. For user provided weights, we used the normalized defuzzified *weight vector* of <0.17, 0.31, 0.52> from the *Linguistic Terms* of "low", "medium" and "high" for Hospital Name, Hospital Address and City. Fig. 4 shows that the proposed solution with user provided and default weights found 2669 and 2701 *Matches*, which were in between 2207 and 2740, the number of *Matches* found by *recordlinkage* and *splink*. It also displays that the proposed solution found 6862 and 6882 *Possible Matches* in the two cases.

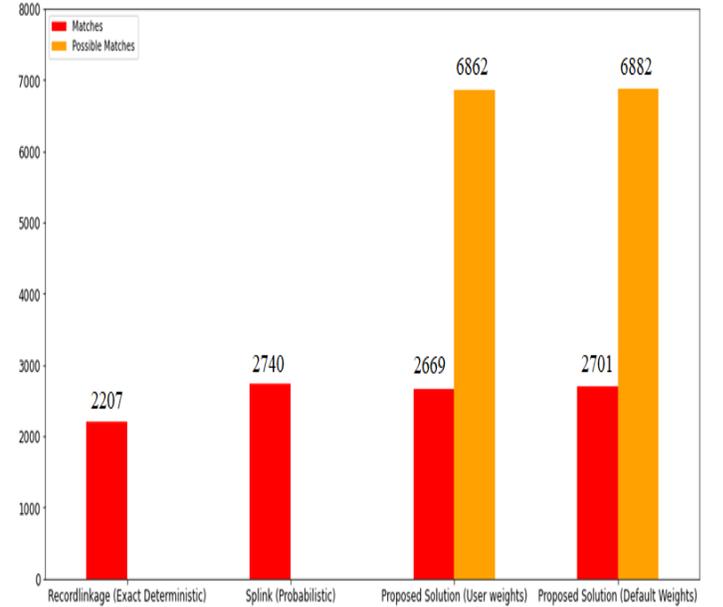

Fig. 4: Comparison with Existing Packages (Crisp Constraint)

Fig. 5 compares the performances of the proposed method with those of *deterministic* and *probabilistic* linkages on the *fuzzily* constrained search space consisting of **998,860** linked record pairs. We implemented *deterministic linkages* by using *exact* string matching (on all *Link Columns*), "Boolean" *Logic Type*, "crisp" *Linkage Type* with the default weight of 1 for each linked column and estimated the *total linkage score* (*TS*) for each record pair as the average of the *match scores* of the *Link Columns*. For *probabilistic linkages*, we calculated the *m*- and *u-probabilities* from a small sample of the constrained search space and then computed a *probabilistic weight vector* for the *Link Columns* using these probabilities. Subsequently, we estimated the *total linkage score* (*TS*) for each record pair as the *dot product* of the *probabilistic weight vector* and the *raw match scores* from the string matches for that pair. For *proposed crisp linkages* (using "Fuzzy" *Logic Type* and "crisp" *Linkage Type*), we estimated the *total linkage score* (*TS*) of each record pair using steps ii. a. & ii. b. of the *Linkage Score Estimation* algorithm (Section IV D); while for *proposed fuzzy linkages* (using "Fuzzy" *Logic Type* and "fuzzy" *Linkage Type*), we estimated the same using steps iii. a. – iii. e. of the same algorithm. Fig. 5 shows that the proposed solution found 9367 & 2459 *Matches*, which were in between 2207 and 99841, the number of *Matches* found by *deterministic* and *probabilistic linkages* respectively. Besides, it also displays that the proposed solution found 392792 & 8460 *Possible Matches* in the two cases.

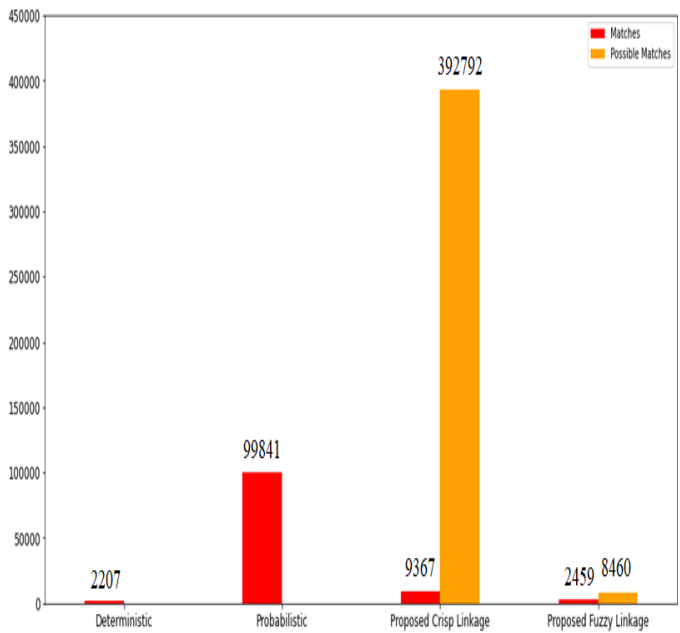

Fig. 5: Comparison with Existing Strategies (Fuzzy Constraint)

Fig. 6 shows the results of proposed *fuzzy linkages* of two record pairs from Hospital Accounts and Hospital Reimbursements (i.e., the two tables used in this experiment) around the boundary between the two clusters, i.e., *Matches* and *Possible Matches*, based on the *Link Columns*, Hospital Name, Hospital Address and City in the order of importance: "low", "medium" and "high". The record pair at the top represents a member assigned to the cluster of *Matches* with the *lowest* estimated *total linkage score* (*TS*), while the pair at the bottom represents a member assigned to the cluster of *Possible Matches* with the *highest* estimated *total linkage score* (*TS*). Note that the record pair at the top matches *fully* on City, Hospital Address and *partially* on Hospital Name. The record pair at the bottom also has *full* matches on City and Hospital Address and only a *partial* match on Hospital Name. However, the record pair at the top matches on *more* words than the pair at the bottom (3 against 1, with the name on the right being fully included in the name on the left in the top pair) on the Hospital Name, which has the lowest importance. Hence, the two record pairs have eventually ended up in two different clusters at their boundary.

| Match | | | |
|---|---|---|---|
| Facility Name | THE HOSPITALS OF PROVIDENCE MEMORIAL CAMPUS | Provider Name | PROVIDENCE MEMORIAL HOSPITAL |
| Address | 2001 N OREGON ST | Provider Street Address | 2001 N OREGON ST |
| City | EL PASO | Provider City | EL PASO |
| State | TX | Provider State | TX |
| ZIP Code | 79902 | Provider Zip Code | 79902 |
| County Name | EL PASO | Total Discharges | 39 |
| Phone Number | (915) 577-6011 | Average Covered Charges | 51544.7 |
| Hospital Type | Acute Care Hospitals | Average Total Payments | 6954.26 |
| Hospital Ownership | Proprietary | Average Medicare Payments | 6050.74 |
| Possible Match | | | |
| Facility Name | ADVENTHEALTH WATERMAN | Provider Name | FLORIDA HOSPITAL WATERMAN |
| Address | 1000 WATERMAN WAY | Provider Street Address | 1000 WATERMAN WAY |
| City | TAVARES | Provider City | TAVARES |
| State | FL | Provider State | FL |
| ZIP Code | 32778 | Provider Zip Code | 32778 |
| County Name | LAKE | Total Discharges | 152 |
| Phone Number | (352) 253-3300 | Average Covered Charges | 24835.8 |
| Hospital Type | Acute Care Hospitals | Average Total Payments | 5596.34 |
| Hospital Ownership | Voluntary non-profit - Church | Average Medicare Payments | 4665.42 |
| Hospital Accounts | | Hospital Reimbursements | |

Fig. 6: Match and Possible Match of record pairs between Hospital Accounts & Hospital Reimbursements

## VI. CONCLUSION

In this paper, we have presented a fuzzy record linkage technique to address some of the challenges associated with deterministic and probabilistic record linkages, with regards to their low true positives or high false positives (false negatives), under uncertainty with imprecise information. We have combined *fuzzy string matching, fuzzy set theory, fuzzy logic, fuzzy constraints, fuzzy weights, fuzzy rules* and *inferences* with *fuzzy clustering* to generate clusters of *Matches* and *Possible Matches* of linked record pairs, to provide a solution that has several advantages over the existing ones. Firstly, it provides a highly parameterized, flexible solution with a wide range of choices. Secondly, it combines multiple strategies to benefit from their complementary advantages. Thirdly, it produces a more balanced, controllable and an objective solution to the desired level of granularity. The number of *Matches* can be expected or controlled to be between those of deterministic and probabilistic linkages to overcome shortcomings associated with either low true positives or high false positives. Lastly, it can serve multiple purposes with results of both *Matches* and *Possible Matches*, thereby better accommodating user interests. For example, a service provider using door-to-door marketing campaigns can target customers linked as *Matches*, whereas the same provider can target the *Possible Matches* for web or email based bulk promotions. It may be noted here that probabilistic linkage can also generate non-overlapping intervals of *Matches* and *Possible Matches*, using multiple (2) crisp cut-off probability thresholds. However, such classification would then become very subjective in nature, based purely on user's choice of the thresholds. Finally, we have established the effectiveness of the proposed record linkage solution through extensive evaluations and performance comparisons.

Currently efforts are underway to link and unify Verizon customers and prospects, associated with multiple disparate data sources, with global identification using the proposed technique.

CONFLICT OF INTEREST STATEMENT

The author states that he is an employee of *Verizon* and this paper addresses work performed in course of the author's employment. There is also a Verizon filed US Patent application, currently pending on this invention.